\documentclass{aa}
\bibpunct{(}{)}{;}{a}{}{,} 

\usepackage{amssymb}
\usepackage[colorlinks=true, urlcolor=blue, linkcolor=red]{hyperref}

\newcommand{\be}{\begin{equation}}
\newcommand{\ee}{\end{equation}}

\begin{document}

\title{The 4m International Liquid Mirror Telescope: Construction, operation, and science}


 \author{J. Surdej\inst{1, 2} \and P. Hickson\inst{3, 4} \and K. Misra\inst{2} \and D. Banerjee\inst{2} \and B. Ailawadhi\inst{2, 5} \and T. Akhunov\inst{6, 7} \and E. Borra\inst{8} \and \\ M. Dubey\inst{2, 9} \and N. Dukiya\inst{2, 9} \and S. Filali\inst{1} \and J. Hellemeier\inst{10} \and M. Kharayat\inst{2} \and B. Kumar\inst{11} \and 
H. Kumar\inst{2} \and M. Kumar\inst{2} \and T.S. Kumar\inst{2} \and 
P. Kumari\inst{2} \and V. Negi\inst{12} \and A. Pospieszalska-Surdej\inst{1} \and S. Prabhavu\inst{2} \and B. Pradhan\inst{1, 13}\and \\ K. Pranshu\inst{2, 14} \and H. Rawat \inst{2} \and B.K. Reddy\inst{2} \and A. Sasidharan Pillai\inst{2, 15} \and K. Singh\inst{2} \and S. Tremblay\inst{3} \and S. Turakhia\inst{3} \and S. Vijay\inst{16}
 }

\institute{
Institute of Astrophysics and Geophysics, Li\`{e}ge University, All\'ee du 6 Ao\^{u}t 19c, 4000 Li\`{e}ge, Belgium \and
Aryabhatta Research Institute of Observational Sciences, Manora Peak, Nainital, 263001, Uttarakhand, India \and
Department of Physics and Astronomy, The University of British Columbia, 6224 Agricultural Road, Vancouver, V6T 1Z1, BC, Canada \and
Outer Space Institute, The University of British Columbia, 325-6224 Agricultural Road, Vancouver, V6T 1Z1, BC, Canada \and
Deen Dayal Upadhyay Gorakhpur University, Civil Lines, Gorakhpur, 273009, Uttar Pradesh, India \and
National University of Uzbekistan, Department of Astronomy and Astrophysics, 100174 Tashkent, Uzbekistan \and
Ulugh Beg Astronomical Institute of the Uzbek Academy of Sciences, Astronomicheskaya 33, 100052 Tashkent, Uzbekistan \and
Centre for Optics, Photonics and Lasers, Universit\'e Laval, 2375 rue de la Terrasse, Qu\'ebec, G1V 0A6, Quebec, Canada \and
Mahatma Jyotiba Phule Rohilkhand University, Pilibhit Bypass Road, Bareilly, 243006, Uttar Pradesh, India \and
Research School of Astronomy \& Astrophysics, Australian National University, Mt Stromlo Observatory, Cotter Road, Weston Creek 2611 ACT,  Australia \and
South-Western Institute for Astronomy Research, Yunnan University, Kunming, 650500, Yunnan, P. R. China \and
Inter-University Centre for Astronomy and Astrophysics (IUCAA), Post Bag 4, Ganeshkhind, Pune 411007, India \and 
Indian Space Research Organization, Bengaluru, Karnataka, India \and
University of Calcutta, 87/1 College Street, Kolkata, 700073, India \and
Instituto de Astrofísica, Pontificia Universidad Católica de Chile, Av. Vicuña Mackenna 4860, 7820436 Macul, Santiago, Chile \and Department of Physics, Ashoka University Rai, Sonipat, Haryana-131029, India
}

\date{Received / Accepted }

\abstract{
The International Liquid Mirror Telescope (ILMT) project was motivated by the need for an inexpensive 4 metre diameter optical telescope that could be devoted entirely to astronomical surveys. Its scientific programmes include the detection and study of transients, variable objects, asteroids, comets, space debris and low surface brightness galaxies. To this end, a collaboration was formed between the Institute of Astrophysics and Geophysics (Li\`{e}ge University, Belgium), several Canadian universities (University of British Columbia, Laval University, University of Montreal, University of Toronto, York University, University of Victoria) and the Aryabhatta Research Institute of Observational Sciences (ARIES, India).  After several years of design work in Belgium and construction in India on the ARIES Devasthal site, the telescope saw its first light on 29 April 2022. Its commissioning phase lasted from May 2022 until June 2023 (beginning of the monsoon). The ILMT was inaugurated on 21 March 2023 and has been in regular operation since October 2023. The telescope continuously observes the sky passing at the zenith using the SDSS $g^{\prime}$, $r^{\prime}$, and $i^{\prime}$ filters. This paper describes the ILMT, its operation, performance and shows some initial results. 
}


\maketitle

\titlerunning{The International Liquid Mirror Telescope}
\authorrunning{Surdej et al.}

\section{Introduction} \label{js:Intro}

It has been known since Newton that the surface of a liquid in rotation around a vertical axis takes the shape of a paraboloid, ideal for focusing a beam of parallel light rays to a point. However, it was not until the mid-19$^{th}$ century that \citet{Capocci}, then director of the Naples observatory, revived the idea of using a rotating container filled with mercury as the main mirror of an astronomical telescope. Discussions of the history of liquid mirror telescopes (LMTs) can be found in \citet{Gibson91} and \citet{surdejetal24a}. 

Liquid mirrors are an inexpensive alternative to large conventional mirrors, for applications in which conventional pointing and tracking is not required. Compared with a solid glass mirror that has to be cast, ground and polished, a rotating liquid mirror is much cheaper to manufacture. For a 4 m mirror, the difference in cost is more than a factor of 30. 

Early liquid mirrors \citep{wood1909mercury} suffered from ripples on the surface of the mercury due to the transmission of vibrations and the difficulty of keeping the angular rotation of the mirror constant. The stellar images obtained were of mediocre quality and unstable. In addition, because of the Earth's rotation, the star images could not be kept fixed in the focal plane of the mirror, whose optical axis was strictly vertical. At the time, the technology was not ready to remedy these problems.

The 4m International Liquid Mirror Telescope (ILMT) presented here has benefited greatly from the technological developments made between 1980 and 2000 by the teams of Borra of Laval University (\citealt{borra1982liquid}) and Hickson of the University of British Columbia (UBC, see \citealt{hickson1994ubc}). To overcome vibration and oscillation, they used an air bearing, which has precision-ground surfaces separated by a thin film of pressurised air. These bearings are virtually frictionless and can therefore ensure very smooth rotation. By using a synchronous motor 
driven by a crystal oscillator, they were also able to eliminate the speed variations that had hampered Wood's and previous instruments. In the case of the ILMT, the mirror is mounted on an air bearing, which facilitates smooth rotation, and is driven by a brushless DC motor integrated within the air bearing. The relative variations in the period of rotation of the mirror are of the order of $10^{-6}$, which makes it possible to maintain a near-perfect paraboloidal shape.

In the early 1990s, optical shop tests of a 1.5m liquid mirror demonstrated that its optical quality was limited by diffraction (\citealt{borra1989diffraction,borra1992liquid}). This led to a 2.7m LMT equipped with a CCD detector operating in time-delay integration (TDI) mode in order to compensate for the Earth's rotation, which provided the first long-exposure astronomical images \citep{hickson1994ubc,hicksonetal1994}. 

Early applications of this technology include NASA's Orbital Debris Observatory (NODO), a 3m LMT used for space debris observations \citep{Potter97} and two 3m class LMTs used for lidar observations of the Earth's upper atmosphere \citep{Sica95,Wuerker02}. The most significant LMT built to date is the 6m diameter Large Zenith Telescope \cite[LZT,][]{Hicksonetal07}, which operated until 2016. The LZT was instrumental in obtaining high-resolution measurements of the Earth's sodium layer for adaptive-optics applications \citep{Pfrommer2009,Pfrommer2010,Pfrommer2014}, and was also used for laser guide star tests \citep{Otarola2016,Hellemeier2020}. However, none of these telescopes were equipped with optical correctors specifically designed for TDI observations and none were located at good astronomical sites.

\begin{figure*}
\centering
\includegraphics[width=17cm]{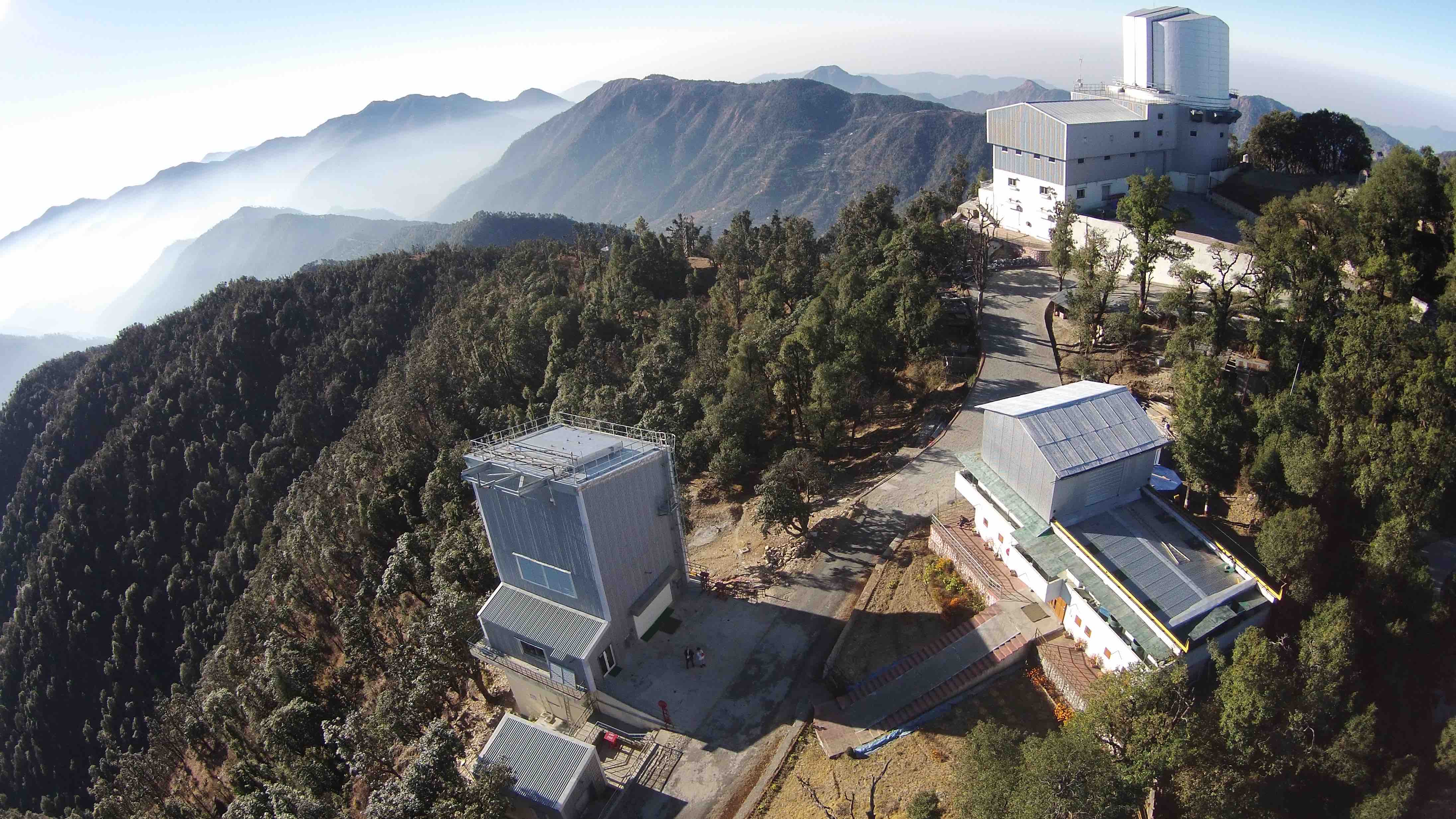} 
\caption{Aerial view of the Devasthal Observatory. The ILMT enclosure is at the bottom left of the image. One easily distinguishes the telescope enclosure, in front of which are the control room and the compressor building. Just to the right is the dome of the 1.3m DFOT and at the top right the dome of the 3.6m DOT.}
\label{js:fig1}
\end{figure*}

The ILMT is located in the foothills of the Himalayas, at the Devasthal Observatory of the Aryabhatta Research Institute of Observational Sciences (ARIES), near Nainital in the northern state of Uttarakhand. Its geographical  coordinates are $79^{\circ}41^{\prime}07.08^{\prime\prime}$ E, $29^{\circ}21^{\prime}41.4^{\prime\prime}$ N, at an altitude of 2,378 metres. Devasthal also hosts the 1.3m Devasthal Fast Optical Telescope (DFOT)  and, since March 2016, the new 3.6m Devasthal Optical Telescope (DOT), India's largest steerable optical telescope (see Fig.~\ref{js:fig1}). 

The main scientific drivers of the ILMT project include:
\begin{enumerate}
\item[i)] the statistical determination of the cosmological parameters $H_0$, $\Omega_M$ and $\Omega_\Lambda$ based on studies of multiply imaged quasars that consist of compact gravitational lens systems, 

\item[ii)] the statistical determination of these same cosmological parameters from supernova surveys, 

\item[iii)] a search for quasars and observational studies of large-scale structure, 

\item[iv)] determination of the trigonometric parallaxes of faint nearby objects (e.g. faint red, white, brown dwarfs, halo stars and other very low-mass stars), 

\item[v)] the detection of high stellar proper motions to probe a new range of small-scale kinematics [stars, trans-Neptunian objects (TNOs), etc.], 

\item[vi)] a wide range of photometric variability studies (cf. photometry of stars, RR Lyrae, novae, supernovae, micro-lensing and other transient events, photometry of variable AGN on timescales ranging from days to years), 

\item[vii)] the detection of low surface brightness and star-forming galaxies, as well as other faint extended objects (galactic nebulae, supernova remnants, etc.), 

\item[viii)] detection and studies of small Solar-System bodies, such as asteroids and comets,


\item[ix)] serendipitous phenomena, and finally, 

\item[x)] the discovery of targets for spectroscopic follow-up studies with the 3.6m DOT. 
\end{enumerate}

More details on the ILMT project and its scientific cases can be found in \citet{claeskensetal01}, \citet{jeanetal2001}, \citet{surdejetal06,Surdej2018}, \citet{kumar2015,kumar2018a,kumar2018b}, \citet{pradhan2018,Pradhanetal19}, and \citet{mandal2020}.



 The present paper is structured as follows. The main components of the ILMT are described in Section~\ref{js:Comp}. Its operation is  discussed in Section~\ref{js:Oper}. The data acquisition, pre-processing and analysis are described in Section~\ref{js:DatAcq} and its current performance in Section~\ref{js:ObsPerf}. Some preliminary scientific results are presented in Section~\ref{js:Sc}. Conclusions form Section~\ref{js:Con}. 

\section{Main components of the ILMT} \label{js:Comp}

The primary components of the ILMT are: 1) a parabolic mirror covered with a thin film of liquid mercury, 2) an air bearing that supports the mirror, and its drive system, 3) the telescope structure, 4) a refractive optical corrector and 5) a CCD camera equipped with broadband filters. These are housed within an enclosure that has a retractable roof hatch.

\subsection{The primary mirror} \label{js:mirror}

In normal operation, the mirror rotates around its vertical axis with a period of approximately 8 seconds, which corresponds to a linear speed at its edge of 1.6 m/s. Its shape is that of a paraboloid with a focal length of 8 m. 

\begin{figure}[h]
\begin{center}
\includegraphics[width=\columnwidth]{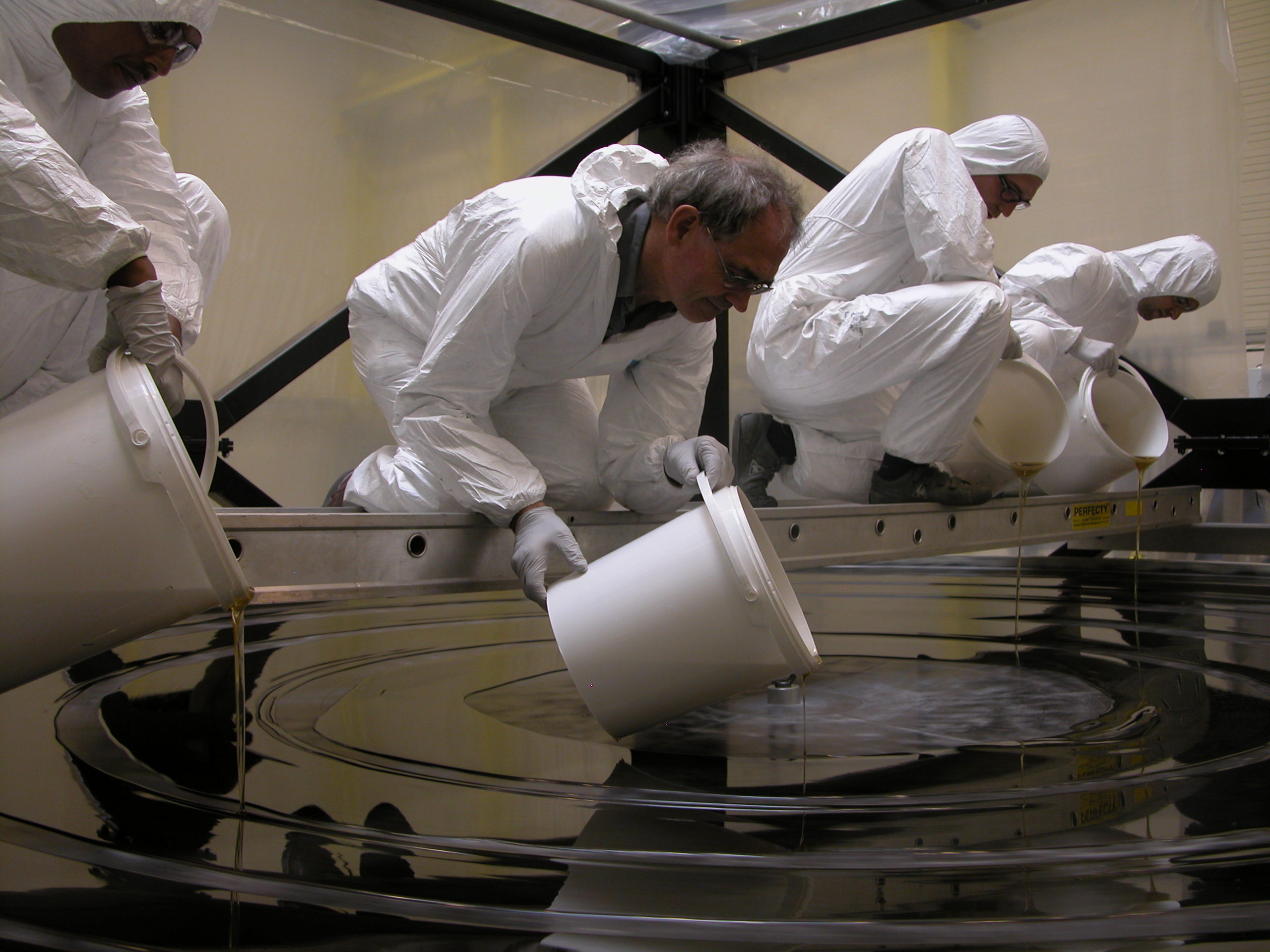} 
\end{center}
\caption{Spincasting of the ILMT mirror. As the mirror rotates around its vertical axis, polyurethane is poured on the surface by members of the ILMT team (from left to right: Brajesh Kumar, Paul Hickson, Fran\c{c}ois Finet and Arnaud Magette). After about half an hour, the resin has solidified.}
\label{js:fig2}
\end{figure}

The design of the mirror aims to minimise weight, while also providing sufficient stiffness to prevent hydrostatic instability. This was achieved using a closed-cell foam core enclosed in a carbon fibre-epoxy skin. The top surface was finished by applying two layers of polyurethane, formed by spincasting (see Fig.~\ref{js:fig2}). This produced a parabolic surface that is accurate to within a few tenths of a millimetre. 

This surface supports a film of mercury that has a thickness of about 3.5 mm. This corresponds to a weight of 650 kg of mercury. The total weight of the rotating mirror is close to one tonne. 

The mirror was designed and built by the company Advanced Mechanical and Optical Systems (AMOS, Li{\`e}ge, Belgium). Before the telescope was transported to India in 2011, several technical difficulties were resolved thanks to critical experiments carried out by UBC and University of Li{\`e}ge astronomers. These included improving the rigidity of the mirror and its mechanical interface with the air bearing, spin casting, and checking the quality of the mercury surface.

\begin{figure}[h]
\begin{center}
\includegraphics[width=\columnwidth]{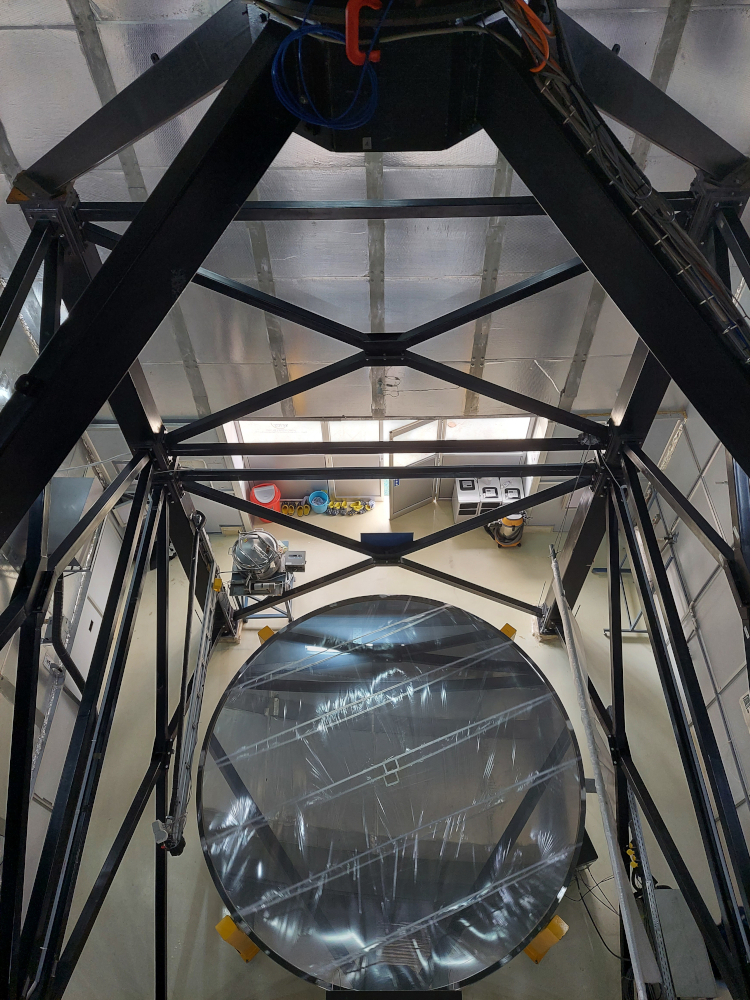} 
\end{center}
\caption{ Top view of the ILMT mirror surfaced  with mercury and covered with thin assembled mylar films, which prevent the formation of ripples on the surface of the mercury. A tube is attached to the aluminium beam, just to the left of the edge of the mirror, to pump the mercury to and from the mirror and the tank visible at the top left side of the mirror. Near the middle left-hand side, one can see one end of a box containing  charcoal to which a black tube is attached. It is used to evacuate the vapour when the mercury has just been transferred to the mirror (see text). }
\label{js:fig3}
\end{figure}

Fig.~\ref{js:fig3} shows the mirror covered with a thin coat of mercury and thin transparent mylar film (2.5 $\mu$m thick and 90 cm wide) to prevent friction between the mercury and the surrounding air that would produce ripples on its surface. At the same time, it helps to confine mercury vapour.

After cutting a small central hole in the mylar film, a tube was installed to extract these vapours and neutralise them with charcoal placed in a rectangular container, visible near the middle left-hand side of Fig.~\ref{js:fig3}. Slightly to the right of this same figure, one can also see the stainless-steel tank into which the mercury is pumped when the rotation of the mirror is stopped for maintenance. 

It is essential to handle mercury safely, as its vapours are hazardous. All possible safety measures are taken during telescope operations. There is no direct contact between mercury and personnel during pumping and cleaning activities. The observatory floor is sealed with epoxy to form a spill container. The mercury is transferred from the sturdy tank to the mirror using a peristaltic pump, which isolates the mercury from the pump mechanism (see Fig.~\ref{js:fig3}). The same figure shows the four yellow safety pillars installed around the ILMT container to prevent it from tipping if the mercury film tears and the mirror becomes unbalanced. The mirror is designed to tilt under such circumstances so as not to damage the air bearing.  Several other pieces of equipment, including a mercury vacuum cleaner, a mercury spill kit, personal protective equipment including masks, gloves, aprons, shoe covers and mercury vapour masks, are used to handle mercury safely. Two mercury vapour detectors continuously monitor the concentration of mercury vapour. 

During observations with the liquid mirror, no mercury vapour is generated after a short time (typically 8 hours) because a thin layer of mercury oxide forms on the surface of the mercury. This transparent layer prevents any further evaporation. 

\subsection{The air bearing and pneumatic systems} \label{js:Air}

The ILMT uses an air-bearing (Kugler: model RT-600T) that is attached to a three-point mount for levelling.  It can support a load of approximately 1,000 kg.

\begin{figure}[h]
\begin{center}
\includegraphics[width=\columnwidth]{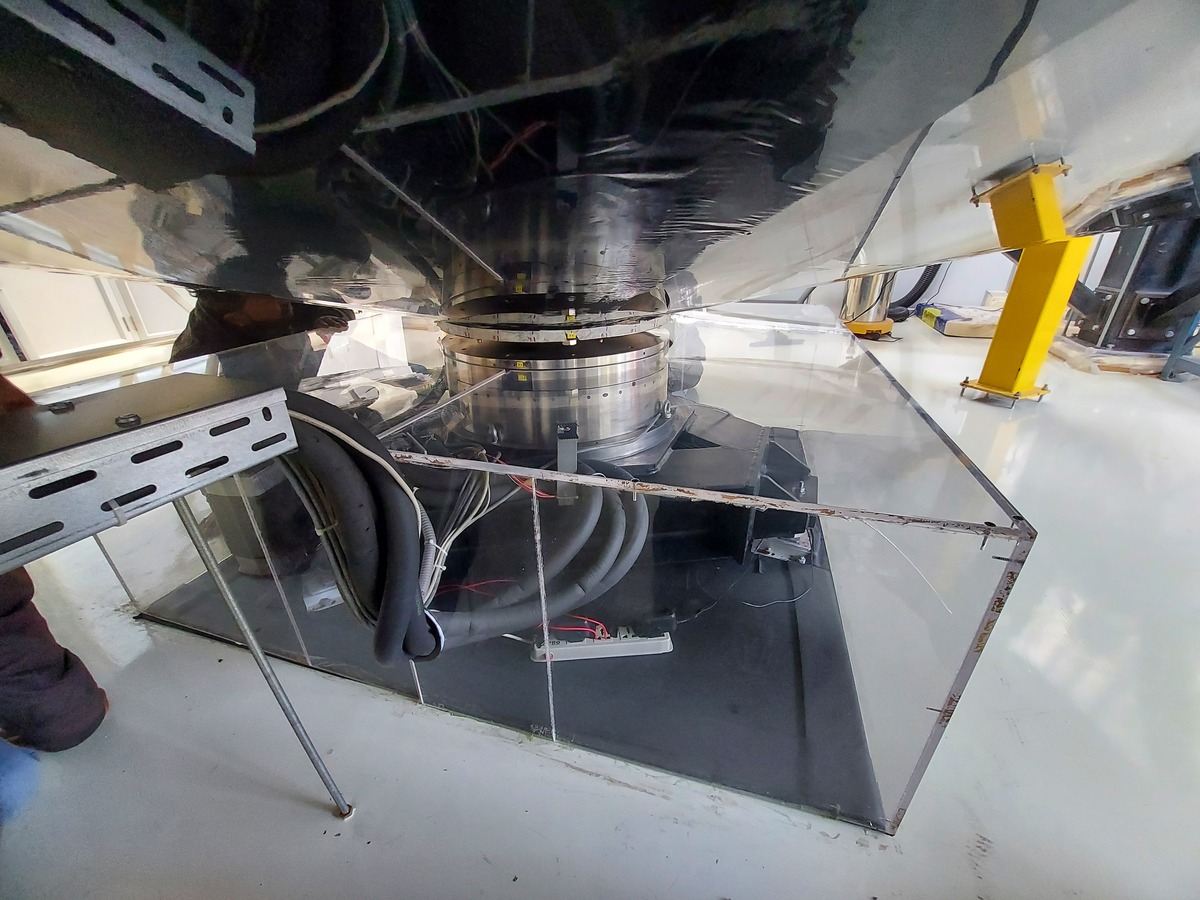} 
\end{center}
\caption{Plexiglass enclosure protecting the air bearing against outside temperature fluctuations, particularly in winter. One of the four yellow safety pillars is seen on this photograph.}
\label{js:fig4}
\end{figure}

In March 2017, the air bearing was precisely positioned just below the centre of the optical corrector (see Section~\ref{js:Cor} below). Its three support pads were firmly anchored to the top surface of the central concrete pier using epoxy (see Fig.~\ref{js:fig4}). The air bearing is enclosed in a Plexiglass box to maintain an ambient temperature of $\sim$ 20$^{\circ}$C. 

\begin{figure}[h]
\begin{center}
\includegraphics[width=\columnwidth]{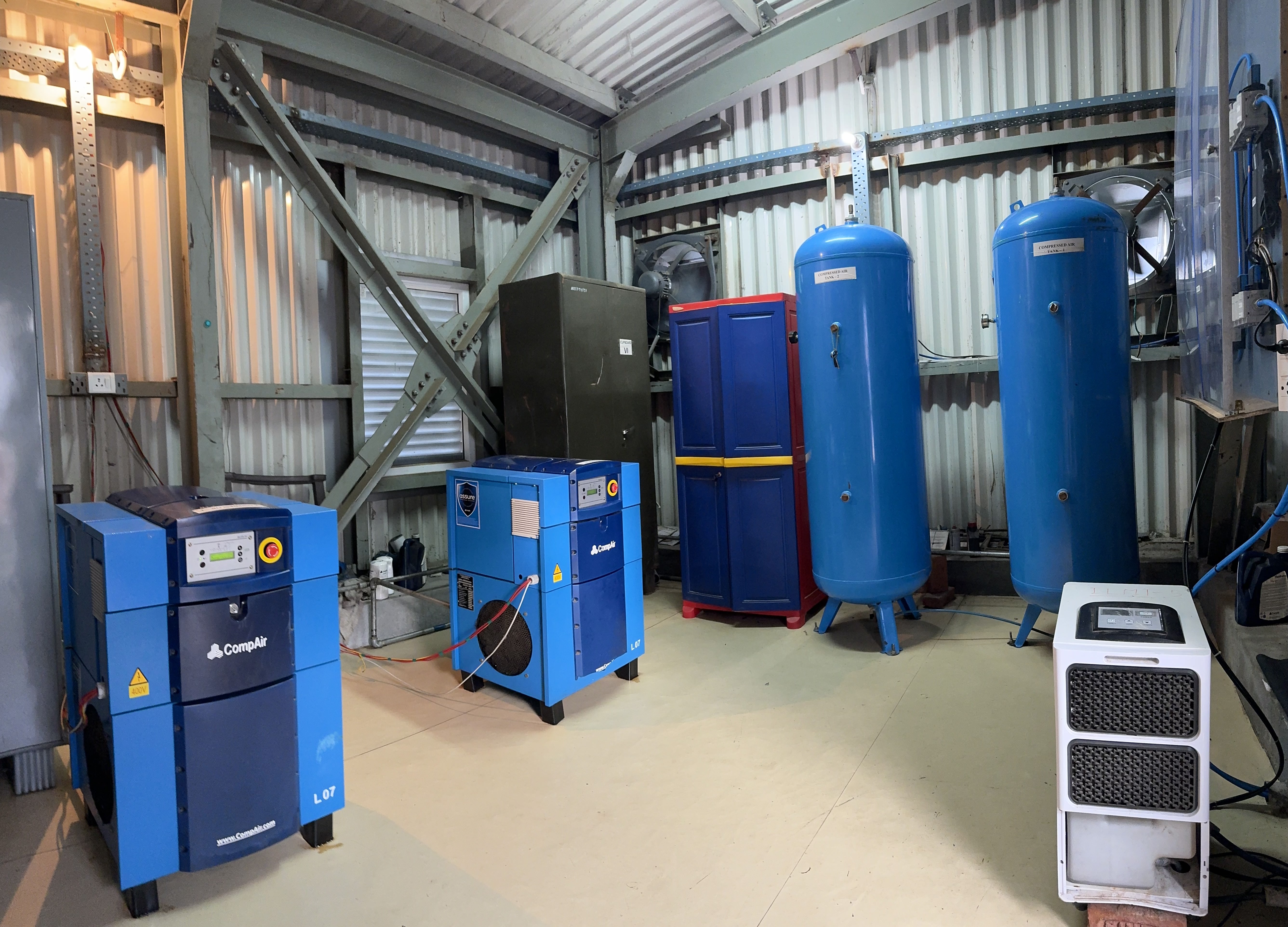} 
\end{center}
\caption{ Two independent and parallel air compressors and tanks are located in the compressor enclosure which is completely separate from the telescope building. Two exhaust fans are used to ventilate the room.}
\label{js:fig5}
\end{figure}

\begin{figure}[h]
\begin{center}
\includegraphics[width=\columnwidth]{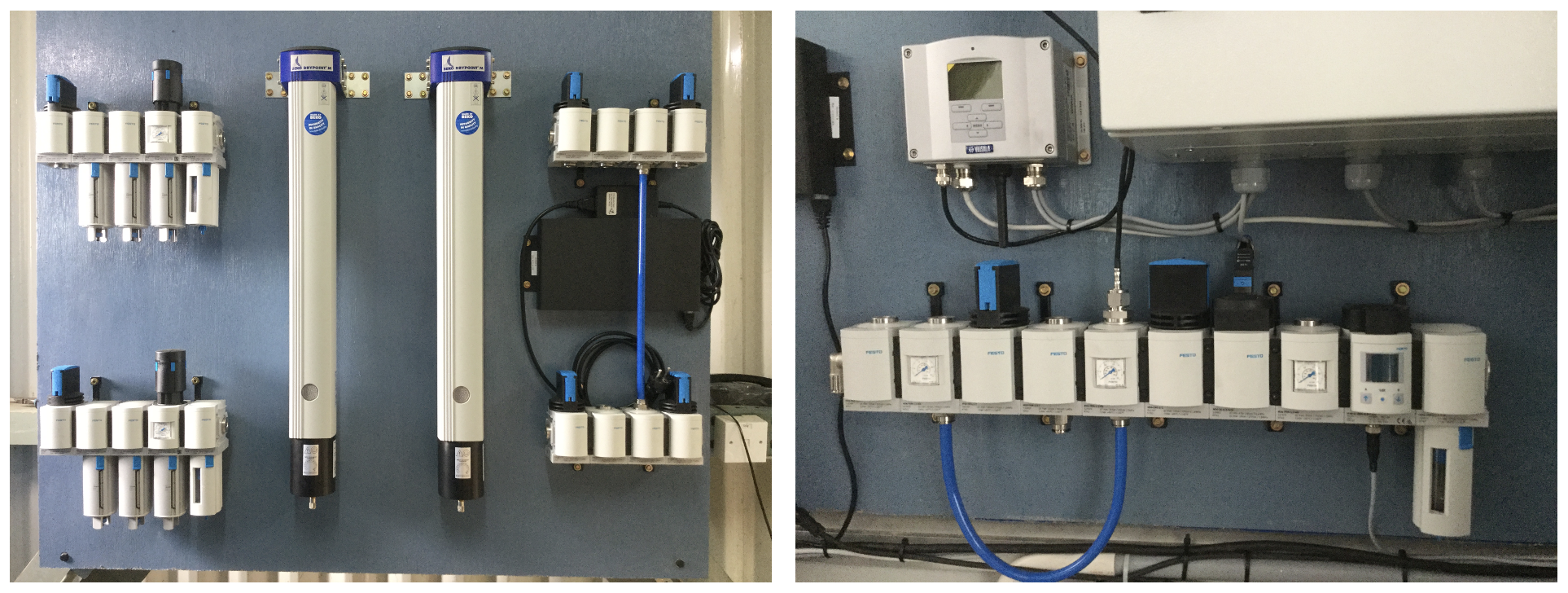} 
\end{center}
\caption{ Double air filtration systems located in the compressor room (left). A separate equipment combines air from each system, providing further filtration, pressure regulation and sensors (right). The latter equipment is located in the telescope control room.}
\label{js:fig6}
\end{figure}

In order to ensure an uninterrupted supply of clean dry air to the air bearing, two independent pneumatic air systems were installed. These consist of two air compressors, two air tanks (see Fig.~\ref{js:fig5}), and two independent filtration systems consisting of coalescing and particulate filters, membrane air dryers and pneumatic valves (see Fig.~\ref{js:fig6}). The system is monitored by pressure, temperature, humidity and dewpoint sensors. If one air system fails or requires maintenance, the second system automatically supplies air to the air bearing. A solenoid valve connected to the dewpoint sensor closes if the dew point exceeds a preset threshold or if the power fails, to prevent any moisture reaching the air bearing.

Observatory Control System (OCS) software \citep{Hickson2019} is used to monitor the status of the compressors, the air system, and the mirror rotation system. It also keeps the telescope in focus, and preprocesses and calibrates the CCD data that are acquired. 

\begin{figure}[h]
\begin{center}
\includegraphics[width=\columnwidth, angle = 270]{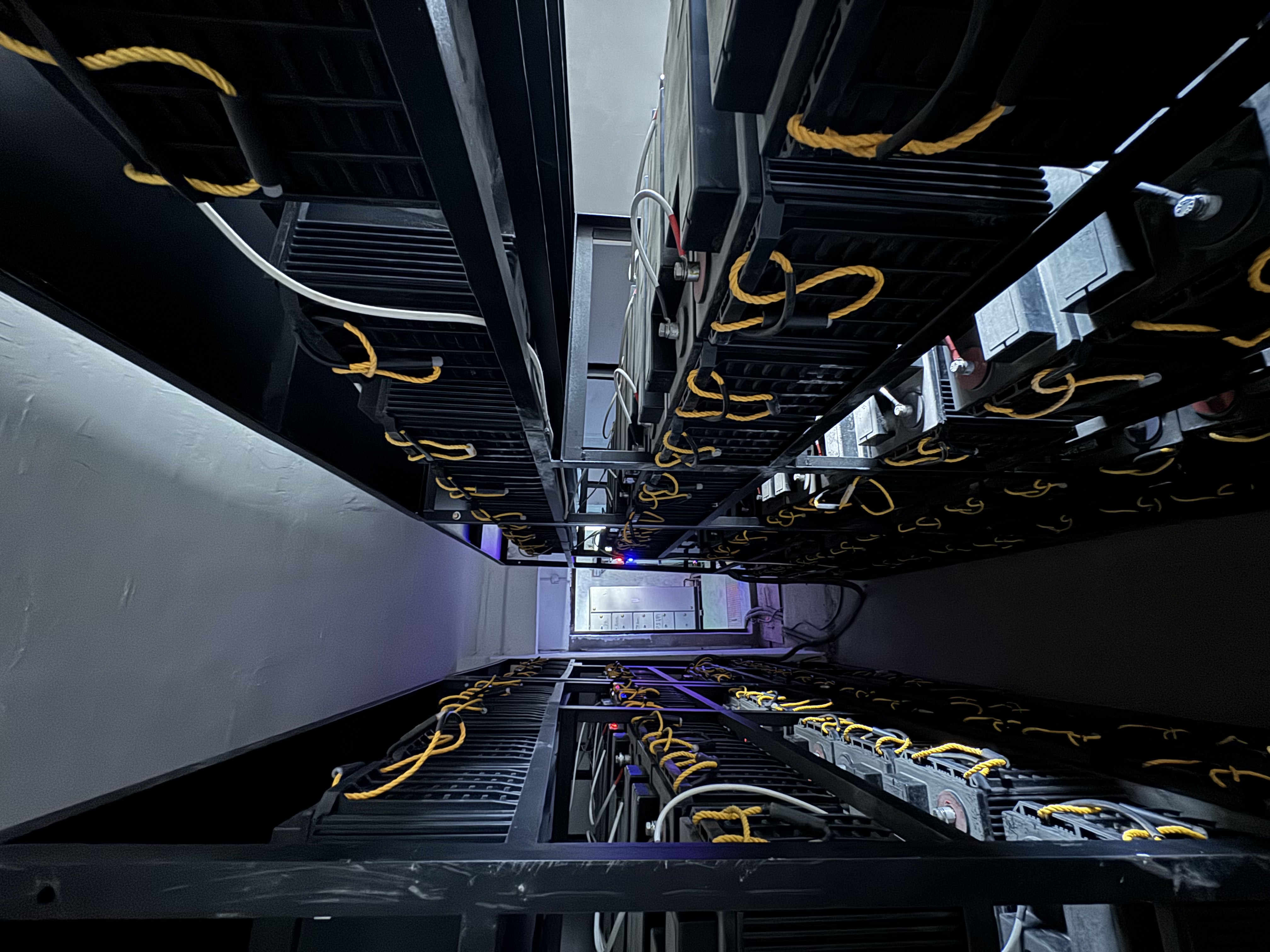} 
\end{center}
\caption{ILMT UPS located in a room below the observatory floor.}
\label{fig7}
\end{figure}

Because of occasional power failures on the main grid, the ILMT facility is equipped with an uninterruptible power supply (UPS) capable of providing the necessary electrical power for at least one hour (see Fig.~\ref{fig7}). The Devasthal observatory is also equipped with two large electrical generators that are switched on if power failures on the grid last for more than 5 minutes.
 
\subsection{Telescope structure and top end} \label{js:Tel}

Because a liquid mirror telescope does not point or track, the telescope structure is greatly simplified and no rotating dome is required. The complex mirror support and tracking systems of conventional telescopes are eliminated. This reduces maintenance and cuts construction and operating costs considerably (\citealt{borra1982liquid,borra1989diffraction,borra2009international,hickson1994ubc,hicksonetal1994}). The main structure of the telescope was built by AMOS.

\begin{figure}[h]
\begin{center}
\includegraphics[width=\columnwidth]{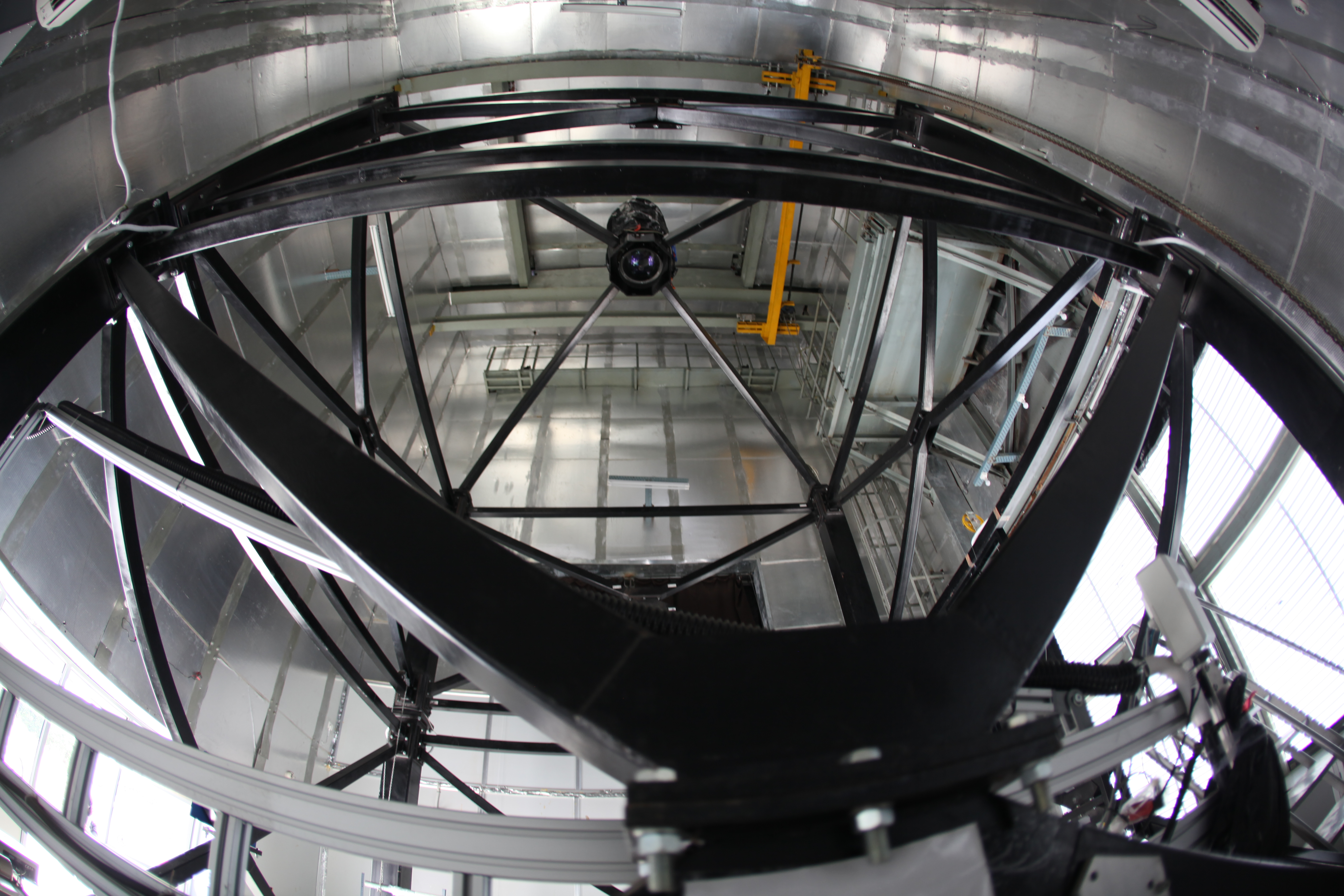} 
\end{center}
\caption{  Telescope structure: the optical corrector is visible at the top.}
\label{fig8}
\end{figure}

Fig.~\ref{fig8} shows the four lower pillars, the lateral reinforcement bars, the four upper pillars and the telescope's spider structure that holds the optical corrector and CCD camera. The top right-hand corner also shows a deployable platform for access to the prime focus of the telescope. 

The central pier, along with four pillars that support the telescope structure, are anchored to bedrock several metres below the floor of the main building. All these elements have been isolated from the outside by the construction of four walls that surround them. This air-conditioned enclosure reduces temperature gradients that could cause differential thermal expansion of the pier and pillars, which can affect the orientation of the mirror rotation axis, and the telescope alignment.

\subsection{The optical corrector} \label{js:Cor}

\begin{figure}[h]
\begin{center}
\includegraphics[width=\columnwidth]{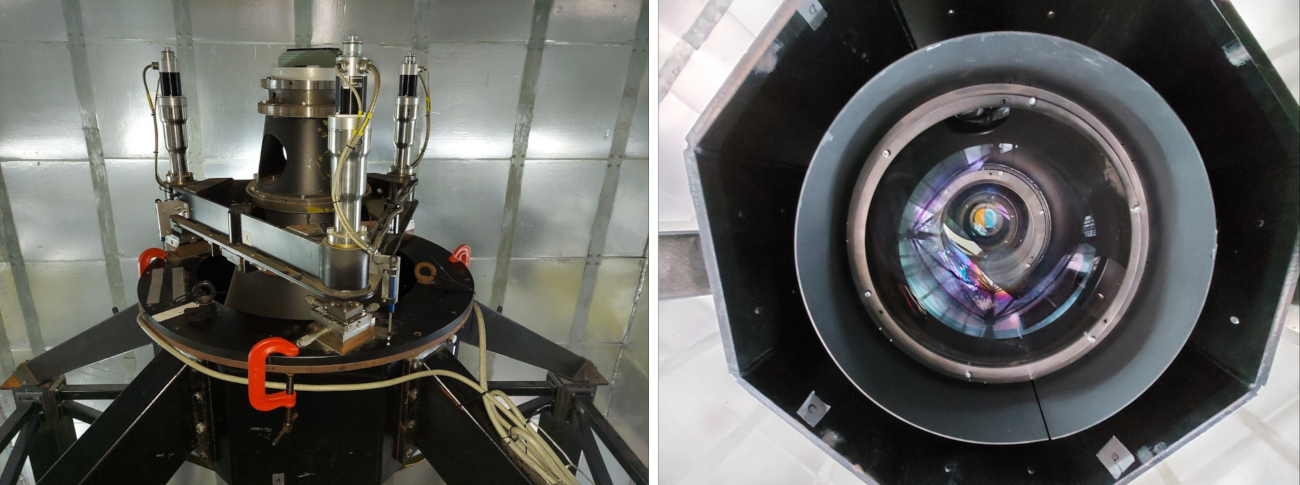} 
\end{center}
\caption{ Views from one side of (left) and from below (right) the optical corrector installed at the prime focus of the ILMT.}
\label{js:fig9}
\end{figure}

Due to the Earth's rotation, the trajectories of stars in the focal plane of a liquid mirror telescope located at the latitude of $+29^\circ 21^{\prime} 41.4^{\prime\prime}$ correspond to hyperbolic trajectories \citep{vangeyte2002study,negietal24,surdejetal24a}. The ILMT is equipped with a five-element optical corrector that is specifically designed for TDI observations (see Fig.~\ref{js:fig9}). It employs tilted and decentred lenses that not only correct the usual off-axis aberrations of the parabolic primary mirror, but also introduce compensating distortion that makes the stellar trajectories linear and equalises the drift rate across the entire field \citep{HickRich98}. The corrector increases the effective focal length of the telescope to 9.434 m, resulting in a focal ratio of 2.36.

Located 7.2~m above the primary mirror, the corrector is supported by an interface that includes tip-tilt and focus actuators (see Fig.~\ref{js:fig9}). Manually operated micrometre screws enable adjustment of its horizontal position. The quality of the observed images is highly dependent on the precision of the centring and levelling of the corrector. The alignment is checked, using defocused star images, before each long period of observation and adjusted if necessary. The focus motion is controlled by the OCS software running on one of the ILMT's computers (ic1). It calculates a temperature correction and automatically adjusts the focus every few seconds.

\subsection{The CCD camera and filters}  \label{js:CCD}

\begin{figure}[h]
\begin{center}
\includegraphics[width=\columnwidth]{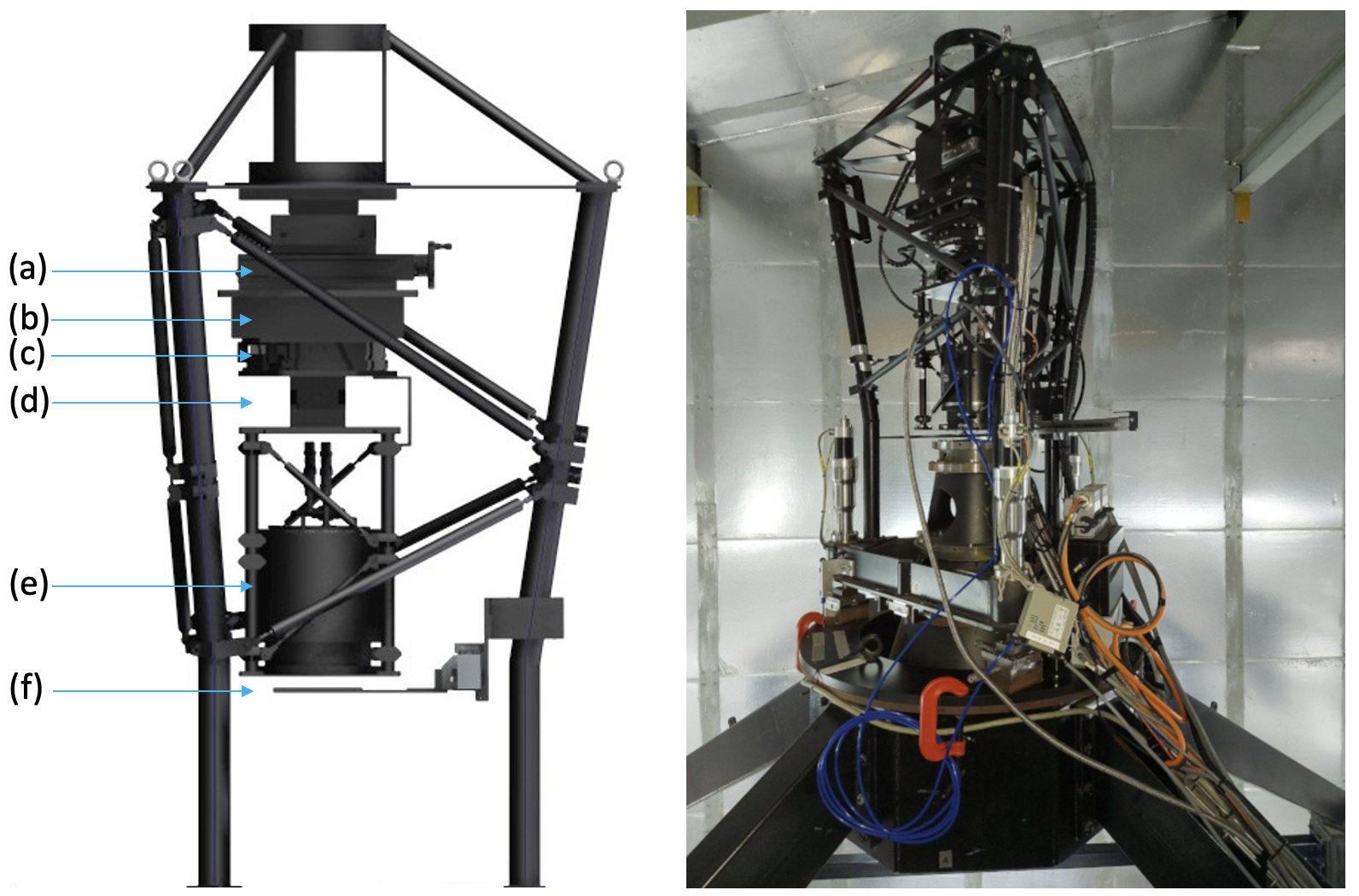} 
\end{center}
\caption{ Schematic of the Socabelec mechanical interface: (a) translation table, (b) tip-tilt table, (c) azimuth rotation table, (d) focus shift, (e) CCD camera and (f) filter tray (left). View of the Socabelec interface installed above the optical corrector at the prime focus of the ILMT (right).}
\label{js:fig10}
\end{figure}

A mechanical structure is located above the optical corrector (see Fig.~\ref{js:fig10}). Designed by CSL (Centre Spatial de Li{\`e}ge, Belgium) and built by Socabelec (Jemeppe-sur-Sambre, Belgium), it holds the CCD camera and a filter tray that contains three broadband filters and a dark slide. It is equipped with mechanisms for adjusting the position of the CCD camera in three axes of translation in addition to rotation about the optical axis. Filter selection, rotation and axial position of the CCD camera are controlled remotely. 

\begin{figure}[h]
\begin{center}
\includegraphics[width=\columnwidth]{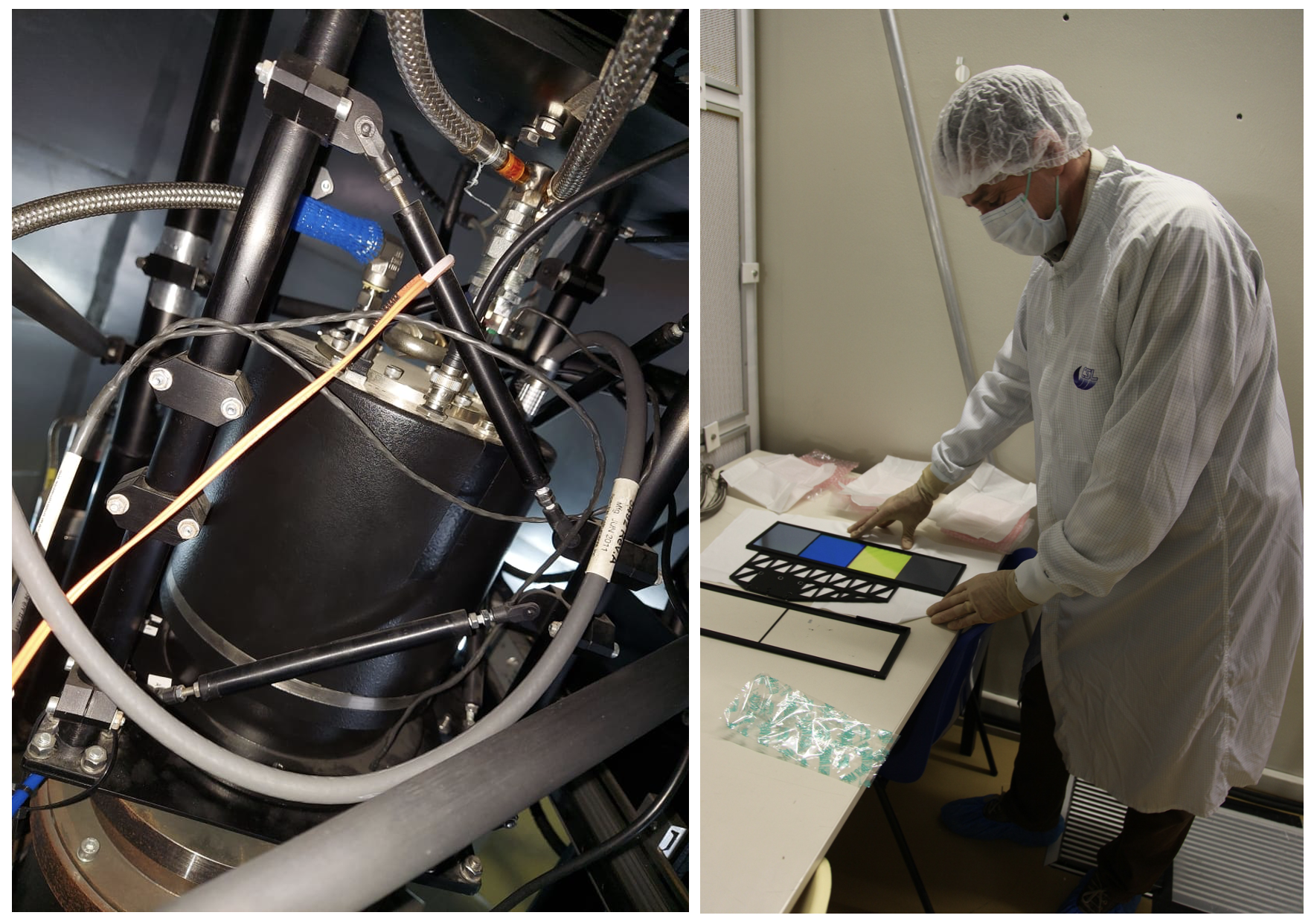} 
\end{center}
\caption{Spectral Instruments CCD camera placed inside the Socabelec mechanical interface at the prime focus of the ILMT (left) and the three SDSS broadband filters installed by Serge Habraken inside the filter tray (right).}
\label{js:fig11}
\end{figure}

The detector is a 4096$\times$4096-pixel CCD camera, having a pixel size of $15 \mu$ (0.3276 arcsec), manufactured by Spectral Instruments (Tucson, Arizona) and operating over the 4000 to 11000 \AA~ spectral range (\citealt{Surdej2018}). The filters (see Fig.~\ref{js:fig11}), which have a typical bandwidth of $\sim 150$ nm and central wavelength $\lambda$, are: $g^{\prime}$ (2 mm GG400 + 3 mm BG38, $\lambda = 468.6$ nm), $r^{\prime}$ (4 mm OG550 + 1 mm BK7, $\lambda = 616.5$ nm), and $i^{\prime}$ (4 mm RG 695 + 1 mm BK7, $\lambda = 748.1$ nm) matching the SDSS photometric system \citep{Fukugita}.

With TDI operation, the CCD detector tracks stars by electronically shifting the relevant charges along the columns of the CCD at the same rate as the images drift, that is the sidereal rate (\citealt{mcgraw1980charge,hall1984faint}). The integration time is equal to the time required for the celestial sources to travel the entire length of the CCD, which is 102.35 s for the ILMT. In this manner, it is possible to record an image whose length in the right ascension direction is limited only by the capacity of the data system \citep{GibHick92}.  The ILMT is thus able to scan a strip of sky centred on the declination of $+29^\circ 21^{\prime} 41.4^{\prime\prime}$. The angular width of the sky strip is 22.3 arcmin, a size limited by that of the CCD (see Section~\ref{js:CCD}). 

\begin{figure*}
\centering
\includegraphics[width=18.5cm]{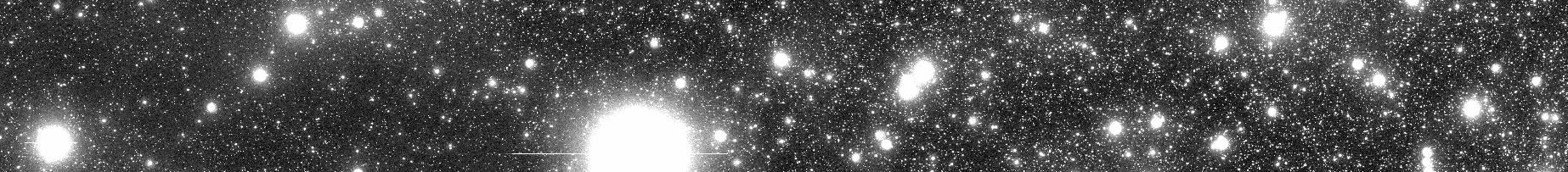} 
\caption{ Example of a single CCD frame recorded in the TDI mode during $\sim$ 17 minutes. This field is centred at the constant declination $\delta = +29^\circ 21^{\prime} 41.4'^{\prime\prime}$ and begins at the right ascension $RA = 5h57m$. North is up, and east is to the left. It covers a field of $22.3^{\prime}$ in $\delta$ and $9 \times 22.3^{\prime}$ in $RA$.}
\label{js:fig12}
\end{figure*}

The total length of time for a TDI scan is limited to about 20 minutes by the camera software. Accordingly, science observations consist of a continuous series of ten 102.35 s scans. The full integration time is reached only after the first 4096 rows of the image have been read, so these rows are discarded. Each image is therefore 
$36864\times4096$ pixels ($200.9 \times 22.3$ arcmin), corresponding to an area of 1.25 square degrees (see Fig.~\ref{js:fig12}). The total sky coverage over a full year is $\sim$ 115 square degrees, with $\sim$ 36 square degrees covered on an average night. This corresponds to a data volume of approximately 15 GB each night.  

The CCD detector is cooled to a temperature close to $-110^\circ$C, by means of a closed-cycle cryo-cooler that uses PT-30 refrigerant, in order to reduce the dark current as much as possible. 

\subsection{The ILMT building } \label{js:Dome}

The ILMT building is located opposite to the 1.3m DFOT building at Devasthal (see Fig.~\ref{js:fig1}) and consists of three parts: the compressor room, the control room and the telescope enclosure. The compressor room is a small building located far enough away from the telescope pier to avoid transmitting vibrations to the mirror. The AMOS telescope control unit, the Socabelec panel which controls the prime-focus interface, the local data server and the mercury monitors are installed in the control room. The control room is isolated from the telescope enclosure and sealed to prevent any leakage of mercury vapour into the room. 

The main enclosure employs a sliding roof shutter that opens to permit observations. The building is equipped with four exhaust fans for ventilation and is air-conditioned during the day. 

Two cameras in the telescope enclosure allow observers in the control room to see the mirror. An all-sky camera and weather station (Vaisala 536 series) are installed on the roof of the telescope enclosure to monitor the sky at night and the weather. 


\section{Operation of the ILMT} \label{js:Oper}

Observations typically begin in October, following the monsoon season \citep{dubeyetal24, surdej2022first}.  System checks are performed for the CCD camera (vacuum pressure, cooling, etc.), the pneumatic system (compressor, valves, etc.), the Socabelec interface mechanisms (filter selection, CCD rotation) and the UPS. The main preparatory activities that take place before starting the mirror are described in the next sub-sections. 

\subsection{Primary mirror and air bearing azimuth alignment}

The mirror is attached to the air bearing by an interface that allows it to tilt if a mercury imbalance occurs. This interface consists of two stainless-steel disks that make contact only in a central area that is precisely ground flat. The lower disk is bolted to the air bearing and the upper disk is bonded to the mirror. The disks are prevented from moving radially by a central pin, but it is possible for them to rotate with respect to each other if enough force is applied. 

It is important that the relative alignment of the two disks be the same as it was during spincasting in order to minimise variations of the surface from a parabola. The correct alignment is indicated by marks on the two disks. If misalignment occurs, it is corrected by lifting the mirror a few millimetres, rotating the  bearing to the correct position, and then lowering the mirror. 

Deviations of the mirror from a paraboloid can be measured by measuring variations in the height (`vertical runout') of the polyurethane surface while the mirror is slowly rotated by hand. This measurement is done using a dial gauge attached to one of the four safety posts that surround the mirror, using a fixture that positions the gauge so that its actuator is in contact with the mirror surface close to the rim. 
The runout deviation is found to be of the order of $\pm 20~\mu$m, which is acceptable.


\subsection{Levelling the mirror}
\label{sec:levelling}

In order for the telescope to produce high-quality images, the axis of rotation of the mirror must be aligned with the apparent vertical direction, that is the vector sum of the gravitational acceleration and the centrifugal acceleration due to the Earth's rotation, ideally to within a fraction of an arcsecond. We refer to this alignment process as `levelling' of the mirror.  This critical activity is repeated whenever the mirror needs to be restarted. 

Levelling is performed using a machinist's precision bubble level (2 arcsec per division) placed on an adjustable triangular base. The bubble level is positioned near the centre of the mirror and the bubble is moved near the centre by adjusting three screws on the triangular base. Next, with air flowing to the bearing, the mirror is slowly rotated by hand, and the position of the bubble is recorded as a function of the rotation angle of the mirror. The three-point mounting screws of the air bearing support are adjusted to minimise variations in the position of the bubble with position angle. The mirror is levelled when the bubble is in the same position for all rotation angles. 

Levelling is currently achieved with a precision better than 1 arcsec. 
{However, variations in the level were observed at different times of the day, likely caused by temperature gradients in the telescope pier.} To address this problem, two improvements were made during the monsoon of 2024. First, a high-precision digital tilt meter was attached to the air bearing base, with its axes aligned with the orientation of the mounting screws, in order to monitor variations in the mirror level. Also, the four telescope pillars and the central pier located 
below the observing floor, which were open to the air, were enclosed with walls, and an air-conditioner installed to regulate the temperature of this area. We expect that this will reduce variations in the mirror level during the coming observations.  


\subsection{Mylar installation and mirror formation}

A transparent polyester film placed above the rotating mirror protects the mercury surface from vortices in the air above the mirror, generated by the rotation. This greatly improves image quality. In addition, the film prevents harmful mercury vapours from spreading outside the mirror. The ILMT uses C-type DuPont Mylar$^\circledR$ polyester film that has a thickness of 2.5 $\mu$m. Several pieces of film are taped together so that they are large enough to cover the entire mirror (see Fig.~\ref{js:fig3}). 

In order to prevent wind from lifting or damaging the film, a support grid is first prepared using  nylon threads. It consists of five equally spaced parallel threads and three  threads in the perpendicular direction. The  film is laid on top of this grid and secured around the entire perimeter of the mirror using adhesive tape. An additional three threads are installed on top of the film. All the threads are also secured with Scotch tape attached to the mirror edge.

\begin{figure*}
\centering
\includegraphics[width=18.5cm]{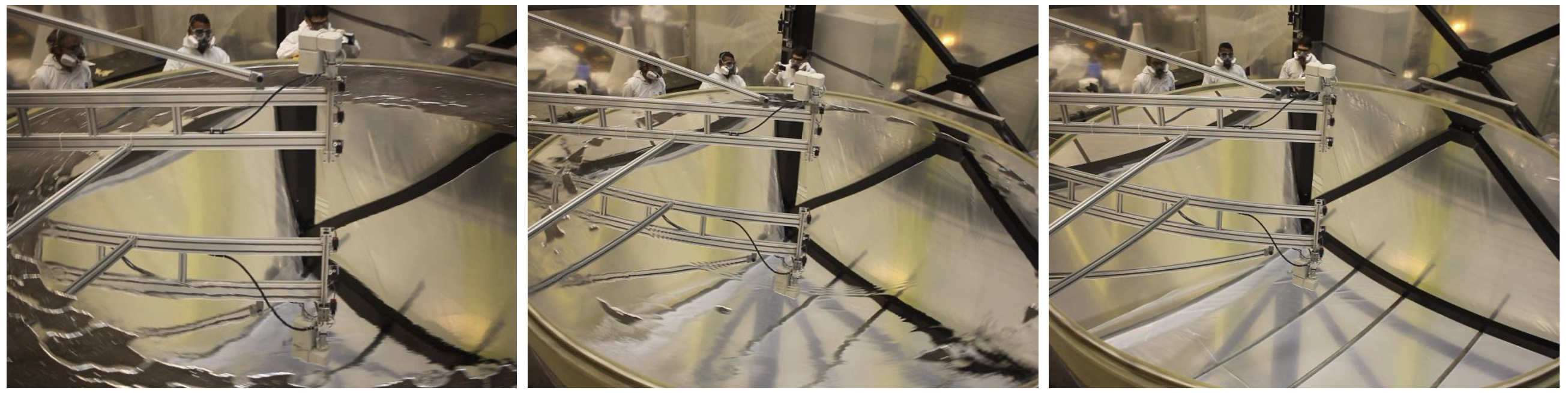} 
\caption{ Three typical phases in the formation of the mercury mirror. Accelerating the rotation first spreads out the mercury (left). However, holes appear and filling them requires the rotation of the mirror to be rapidly decelerated and then accelerated again (centre). When the holes are filled, the control loop is engaged to rotate the mirror at the correct speed (right). These tests were first performed in a secure enclosure at AMOS.}
\label{js:fig13}
\end{figure*}

After the mylar cover is installed, the mirror can be started. The ventilation fans are turned on  and the roof hatch is opened to provide ventilation of the main enclosure before the mercury is poured. Approximately 50 litres of mercury are pumped onto the mirror and the mercury transfer system is retracted.  The mirror is then rotated by the air-bearing motor, according to commands from a technician using a handle paddle. The mirror is successively accelerated and decelerated to fill  holes that form when the mercury spreads out towards the periphery. Once the mercury surface is closed, the rotation is locked to the appropriate speed (see Fig.~\ref{js:fig13}). It takes about two hours for the mercury to completely stabilise. After that, observations can begin. 


\subsection{Final adjustments}

On the first night after the mirror is started, several engineering observations are made in order to optimise the TDI scan rate of the CCD camera. Next, dark images are taken to verify that there are no light leaks. The optimum focus is then determined using star images. The azimuth angle of the CCD camera is adjusted, if necessary, to ensure that the columns of the CCD are aligned as closely as possible with the E-W direction. Finally, defocussed star images are obtained to check the alignment of the optical corrector. If necessary, the horizontal centring of the corrector with respect to the mirror axis is adjusted by means of two micrometres, one each for the E-W and N-S directions.

The focus is returned to the optimal position and the autofocus system is engaged. Science observations can then commence.

During the observing season, the mirror turns continuously for at least several months. The mylar film is periodically cleaned by blowing compressed air over its surface to remove dust and insects. If the mirror stops due to a technical problem, the mercury is pumped into the tank and the mylar cover is removed. The surface of the mirror is then cleaned using isopropanol, and a mercury vacuum cleaner is used to remove any visible mercury droplets. A new mylar cover is installed and the mercury is pumped back onto the mirror. The mirror rotation is then restarted and the mercury surface closed. This whole procedure takes less than 10 hours. 

\section{Data acquisition, pre-processing and analysis} \label{js:DatAcq}

The recorded ILMT data are stored on disk and are then automatically preprocessed by the OCS software to remove instrumental signatures (dark and flat field corrections) and to provide an initial astrometric and photometric calibration. A unique feature of TDI mode observations is that the effects of dark and  sensitivity,  namely the flat field, are resultant effects averaged over all CCD columns (along the scan direction) rather than being pixel-dependent, as in conventional direct imaging. For this reason, both dark and flat-field calibrations are one-dimensional and are characterised by a very high signal-to-noise ratio. They are a function only of positions along the N-S direction as the CCD is read along the E-W direction.


The initial astrometric calibration makes use of  reference stars in the image.  In this step, a simulated image is generated by adding scaled point-spread functions (PSFs) at positions of stars in the {\it Gaia} DR3 catalogue to a blank image of the same size. A cross correlation between the actual and simulated image gives initial $\alpha$ and $\delta$ position offsets. The coordinates of the {\it Gaia} stars are then compared with centroid positions of the corresponding images in order to determine an astrometric solution. The TDI mode ensures that the column direction corresponds exactly to right ascension and the row direction to declination. No rotation matrix is needed. From this calibration, world coordinate system \cite[WCS,][]{Greisen2002} entries are added to the FITS header. The projection employed is Plate Carr\'ee \cite[CAR,][]{Calabretta2002}, which is appropriate for our TDI images. This initial astrometric calibration, which uses only linear terms, results in a typical RMS position accuracy of $\sim$ 0.3 arcsec. 

The initial photometric calibration is performed by PSF fitting to obtain instrumental magnitudes of the reference stars. Magnitudes in the {\it Gaia} $G$, $G_{BP}$ and $G_{RP}$ bands are transformed to the appropriate ILMT band using transformations published by \cite{Jordi2006}. By comparing the ILMT instrumental magnitudes with the transformed magnitudes of the {\it Gaia} stars, a median magnitude zero point is found and added to the FITS header. 

For further analysis, two independent and complementary pipelines have been created, one at ARIES and the other at UBC. The ARIES astrometric pipeline  \citep{dukiya2022astrometric,dukiyaetal24,negietal24,Ailawadhietal24} uses the transformations \begin{align}
\alpha & = f_1 + (y-y_0) * f_2 + (x-x_0) * f_3, \nonumber \\
\delta & = g_1 + (x-x_0) * g_2 + (x-x_0)*(x-x_0)*g_3,
\label{js:eq1}
\end{align}
where $x$ and $y$ refer to the CCD axes aligned along the declination ($\delta$) and right ascension ($\alpha$) directions, respectively, $x_0$, $y_0$ are the centre $x$, $y$ pixels of the TDI CCD frame, and $f_1$, $f_2$, $f_3$, $g_1$, $g_2$, $g_3$ are free parameters.

This second order calibration achieves an accuracy slightly better than 0.1 arcsec. The improvement occurs because the optical corrector introduces asymmetric distortion in the N-S direction, which is largely corrected by the second-order terms in the transformation. 

The ARIES pipeline also includes a photometric calibration in which aperture magnitudes, using an aperture radius equal to twice the full-width at half maximum (FWHM) of stellar images, are calibrated using the Pan-STARRS-1 reference star catalogue  (\citealt{Ailawadhietal24}). The calibrated magnitudes so derived are comparable to those obtained by PSF fitting during the data preprocessing phase.  

\begin{figure}[h]
\begin{center}
\includegraphics[width=\columnwidth]{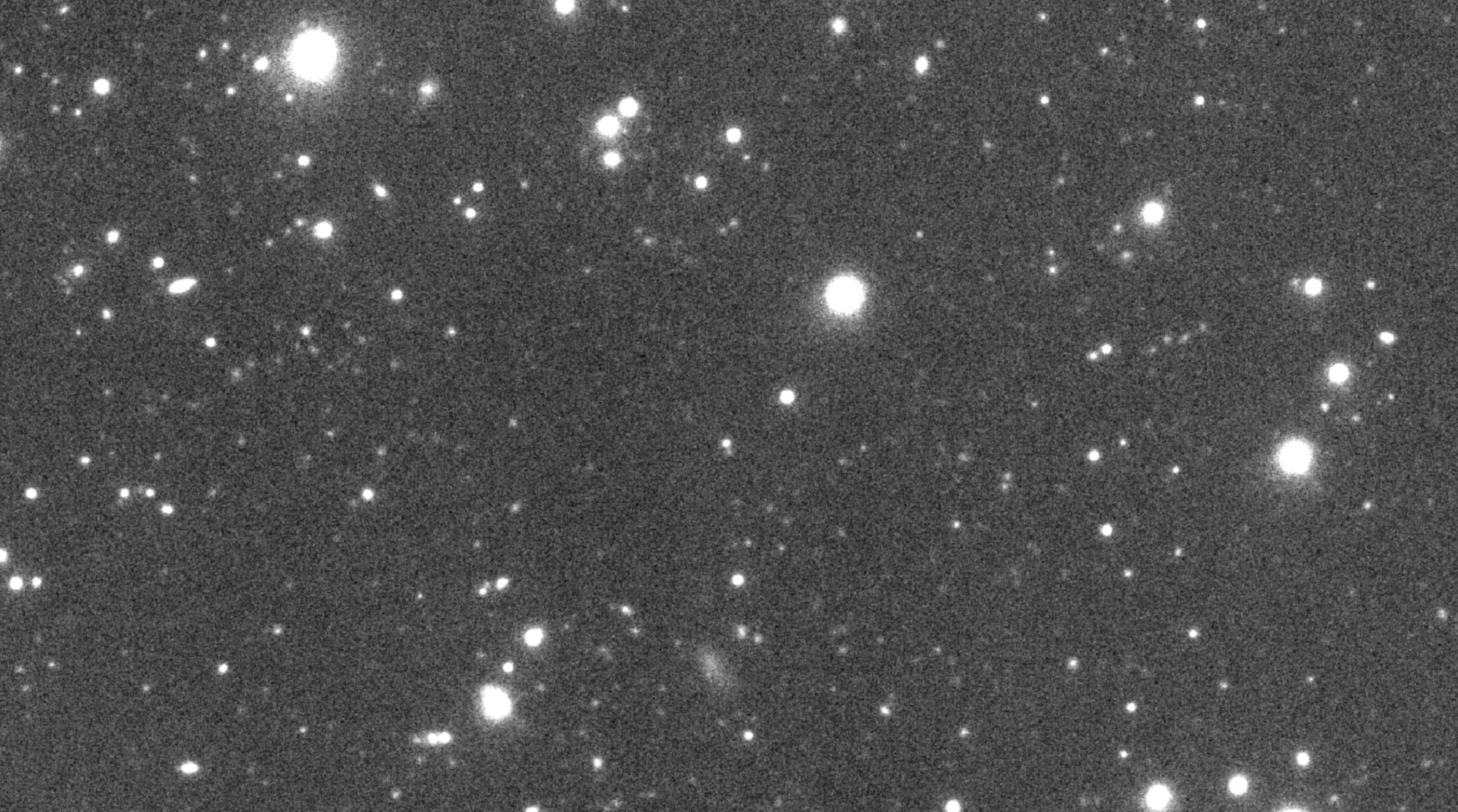} 
\end{center}
\caption{Small portion of an ILMT co-added map. This $i^\prime$-band image was created from 29 individual observations. The region shown is approximately $10 \times 5$ arcmin and is centred at 12h34m51s +29$^\circ 34^{\prime} 50'^{\prime\prime}$. North is up, and east is to the left. One can see numerous faint galaxies.}
\label{fig14}
\end{figure}

The UBC pipeline applies an astrometric correction that uses fixed second and third-order distortion coefficients to obtain an accuracy of $\sim 0.1$ arcsec. From the calibrated images, a series of individual and co-added sky maps are created. These are $8096\times8096$-pixel images in which the coordinate axes are strictly proportional to J2000 right ascension and declination, with a fixed image scale in declination of 0.327 arcsec/pixel. The image scale in right ascension necessarily varies slightly with declination due to the nature of the CAR projection. The maps are centred at a declination of  $\delta = +29^\circ 21^{\prime} 41.4'^{\prime\prime}$ and at intervals of 3:00 minutes of right ascension. Individual maps are created for each night of observations. For a given filter, these can be compared with a complete set of co-added maps that employ all science images obtained with the ILMT in previous observing seasons. Each pixel in a co-added map is the median of the corresponding pixels in all calibrated ILMT images that include that sky position, which can be as many as 30 at the present time. An example of an $i^\prime$-band coadded map is shown in Fig. \ref{fig14}. The limiting magnitude in the co-added maps can be as much as 1.8 magnitude fainter than that of individual ILMT images. 

\begin{figure}[h]
\begin{center}
\includegraphics[width=\columnwidth]{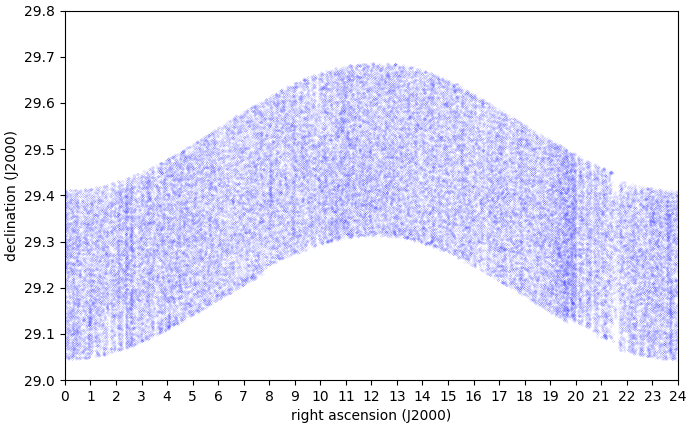} 
\end{center}
\caption{Sky coverage of the ILMT survey, as of 15 June 2024. Each dot represents 200 sources in the preliminary source catalogue.}
\label{fig15}
\end{figure}

\begin{figure}[h]
\begin{center}
\includegraphics[width=\columnwidth]{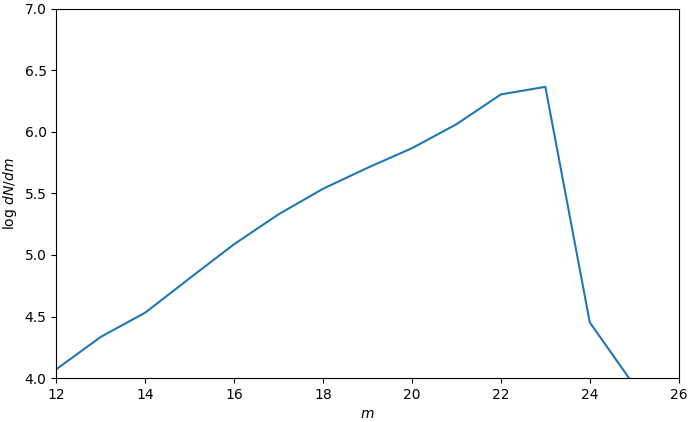} 
\end{center}
\caption{Differential source counts in the $i^\prime$ band. From this we infer that the catalogue is substantially complete to $i^\prime \simeq 23$.}
\label{fig16}
\end{figure}

A preliminary ILMT source catalogue has also been generated at UBC. Objects are detected by matched filtering: the images are first convolved with the PSF and the positions of local intensity maxima are determined. Photometry is then performed by means of PSF fitting. Fig. \ref{fig15} shows the sky coverage of the ILMT survey as of June 2024 and Fig. \ref{fig16} shows differential source counts in the $i^\prime$ band. Further details of the sky maps and source catalogue will be described in a forthcoming paper (Hickson et al. in preparation). 

\begin{figure}[h]
\begin{center}
\includegraphics[width=\columnwidth]{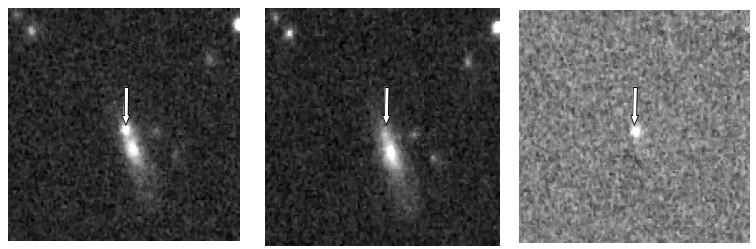} 
\end{center}
\caption{ILMT detection of the supernova 2024cjb in the $i^{\prime}$-band. The left image shows the detection on 16/2/2024, the middle image was recorded on 9/2/2024 and represents  the reference image subtracted from the previous one and the right image corresponds to the subtraction between the left and middle images. We see that the host galaxy is very well subtracted. The white arrow indicates the location of SN 2024cjb at 9h11m27.5s +29$^\circ 29^{\prime} 36^{\prime\prime}$.
The size of the field is $1^{\prime} \times 1^{\prime}$. North is up, and east is to the left.}
\label{js:fig17}
\end{figure}

The ILMT is well-suited to detecting and monitoring optical transients, as nearly the same region of sky is observed each night. As each TDI image contains a very large number of sources, an automated pipeline is essential for detecting transients. Such a pipeline, which employs image subtraction and an automated transient detection and classification system based on a convolutional neural network, was developed at ARIES \citep{pranshuetal24,pranshuetal25}. It enables photometrically and/or astrometrically variable objects to be identified, including unresolved objects superimposed on extended ones (e.g. multiply imaged quasars, supernovae, etc.). One example of such detections is given in Fig.~\ref{js:fig17}.

\section{Performance of the ILMT}\label{js:ObsPerf} 

How does the performance of the ILMT compare to that of a conventional telescope, at the same location, observing  the zenith? The essential difference is in the primary mirror. Mercury has a reflectivity of about 76\%, which is lower than that of aluminium ($\sim 90$\%) and silver ($\sim 95$\%). However, the reflectivity of conventional mirrors degrades with time. Even with periodic cleaning, the reflectivity of protected aluminium decreases to about 84\% after 31 months \citep{Magrath1997}. In contrast, the full reflectivity of mercury is achieved every time the mirror is restarted and there is little degradation over the typical operating period. 

The mylar film that covers the mirror prevents surface waves that would greatly degrade the quality of the images. However, the film does have some undesirable impacts on performance. Interferometric tests, at a wavelength of 633 nm, indicate that this film induces an RMS wavefront error of $\sigma = 0.031$ waves (19.9 nm) in the double-pass configuration used by the ILMT. A consequence of this is that a fraction of the light from a stellar source is diffracted into an extended halo, removing light from the core of the PSF. The fraction lost can be estimated using the Mar\'echal approximation \citep{Marechal1947} and is
\be
  1-S \simeq 1-\exp(-k^2\sigma^2),
\ee
where $S$ is the Strehl ratio and $k = 2\pi/\lambda$ is the optical wave number. For the wavelength bands of the ILMT, this fraction ranges from 2.7\% for the $i^\prime$ band to 6.9\% for the $g^\prime$ band. 

A second source of light loss is dielectric reflection from the mylar film. The refractive index of the film is 1.65, which results in 6.0\% of the light being lost by reflection at each surface. In double pass, this removes 21.9\% of the light. This light is reflected out of the beam and does not contribute to the PSF. 

Combining these losses we find that the liquid mirror has a throughput of about 55 -- 58\%, compared to 87\% for a typical aluminised glass mirror. The sensitivity penalty is $\Delta m \simeq 0.40$ in the limiting magnitude of the telescope. 

The image quality of the ILMT is limited primarily by the natural seeing at the observatory site. The median seeing at 0.5 $\mu$m is estimated to be $\sim 1.2$ arcsec \citep{sagar2000evaluation}. 

During the first two or three hours of the night, the ILMT image quality is noticeably worse, of the order of 2 -- 3 arcsec due to local turbulence within the enclosure, a result of the mirror and telescope structure being several degrees warmer than the outside air. Once the temperature differential has fallen to about $1^\circ$ C or less, the image quality stabilises, with a FWHM that is typically between 1 and 2 arcsec {(see Fig.~\ref{fig18}). As an example, on 10 October 2024, the dome could not be ventilated at the beginning of the night because of high humidity. Fig.~\ref{fig18} shows that the seeing improves during the night. Ideally, improvements to the air conditioning of the ILMT enclosure should eliminate such dome seeing effects. 

We also see from the seeing measurements reported in Fig.~\ref{fig18} that the image quality does not vary significantly with declination, indicating that the TDI optical corrector corrects the TDI distortion quite well.

\begin{figure}[h]
\begin{center}
\includegraphics[width=\columnwidth]{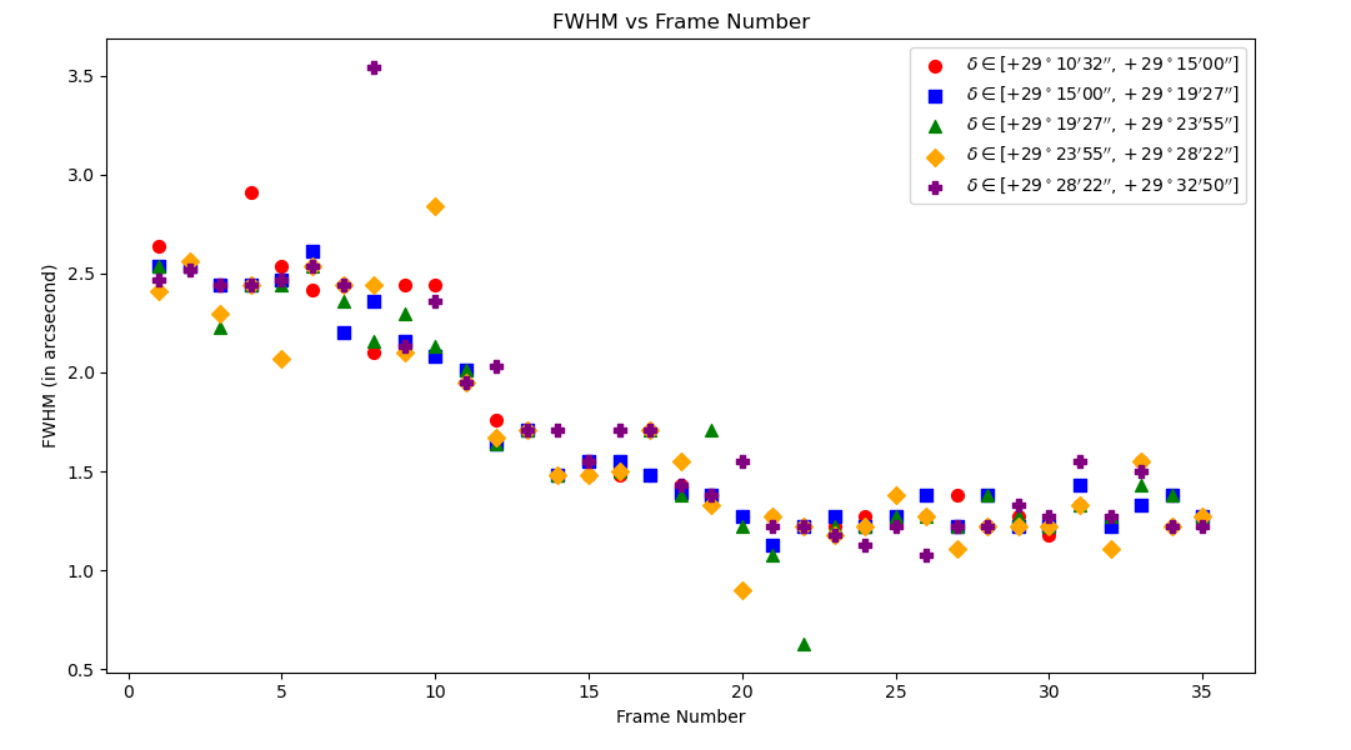} 
\end{center}
\caption{Seeing measurements during a typical night. Each measurement refers to the average full width at half maximum (FWHM) expressed in arcsec of the stellar images recorded over a long CCD frame acquired during typically 17 minutes. During this night of 10 October 2024, 35 such frames have been acquired. The different symbols and colours refer to five different ranges of declination.} 
\label{fig18}
\end{figure}

From initial observations conducted with the ILMT, the following limiting magnitudes were determined for individual exposures: 21.9, 21.7, and 21.5 in the $g^{\prime}$,  $r^{\prime}$, and  $i^{\prime}$-bands, respectively, for a 3-$\sigma$ detection using an exposure time of 102 s (\citealt{Ailawadhietal24}). Let us however note that under optimal atmospheric conditions (seeing better than 1.5 arcsec, good transparency and dark sky), objects fainter than mag = 22 are detected in the g', r', and i' spectral bands.

\section{Science with the ILMT} \label{js:Sc}

Some research programmes (e.g. long-term surveys and photometric monitoring programmes) can be impractical with conventional telescopes but become possible with survey telescopes such as the ILMT. This is particularly true for types of research that do not require the observation of a specific region of the sky. The ILMT is primarily dedicated to projects of high scientific interest dealing with astrometric and photometric variability. 

Preliminary scientific results obtained with the ILMT include the detection and characterisation of solar system objects (\citealt{pospieetal24}), low surface brightness galaxies (\citealt{fuetal24}), galactic (\citealt{grewaletal24}) and extra-galactic objects (\citealt{Akhunovetal24,sunetal24,kumaretal24}).  

An unexpected application of the ILMT is the detection and characterisation of Earth satellites and space debris. An examination of images obtained during the first observing season revealed 301 tracks left by such objects. About 2/3 were identified with catalogued satellites and known debris. The remaining objects constitute an unknown (or unpublished) population. Their magnitudes range from $V = 6.4$ to 19.5 and the estimated altitudes range from near-Earth orbit to cislunar distances \citep{hicksonetal24a,hickson2024serendipitous}. 

\subsection{Data availability}

As a first step towards disseminating the ILMT survey data to the scientific community, access to the data is provided via the \href{https://cloud.aries.res.in/index.php/s/xPER9Y3XuaCsTL9}{ARIESCloud service} \citep{misraet24}. Within the main folder `ILMT Zenithal Survey Data', there are two sub-folders entitled `Observing run - 1' and `Observing run - 2'. Each of these contains three sub-folders, namely `calibration\_files', `rawdata' and `wcs corrected data', as well as a Readme.txt file. The best 1-D dark frames and flat fields in the $g^{\prime}$, $r^{\prime}$, and $i^{\prime}$ bands are accessible in the  `calibration\_files' folder. If necessary, they can be used to preprocess the raw data again. The raw data and wcs corrected data folders contain the raw, preprocessed and astrometrically corrected files, respectively. The data contained in these two folders is sorted by the date on which the observations were acquired with the telescope. The naming convention used for the raw and astrometrically calibrated files are RAW\_yyyymmdd\_filter\_LST.fits and yyyymmdd\_filter\_LST.fits, respectively. For example, an image recorded on 1 November 2024 at 13:45 LST with the $r^{\prime}$ filter is named RAW\_20241101\_r\_13h45m.fits and the corresponding astrometrically corrected file is named 20241101\_r\_13h45m.fits. LST stands for the local sidereal time at the start of the observation. It corresponds to the right ascension of the first celestial objects recorded on the CCD frame.

More details on data policy and access can be found on the \href{https://www.aries.res.in/facilities/astronomical-telescopes/ilmt}{ARIES web site}. The last data release took place in December 2024 and the next one is planned for December 2025.

\section{Conclusions} \label{js:Con}

The ILMT is the first liquid mirror telescope dedicated to astronomical surveys. As the Earth rotates and the seasons change, the telescope scans the same strip (with a daily 4 minute shift in right ascension) of constant current-epoch declination, equal to the latitude of the observatory. The strip of sky thus samples a fair slice of the Universe. The ILMT field of view passes very close to the north galactic pole ($\delta$(J2000) = +27$^{\circ}08^{\prime}$), making it well-suited for extragalactic studies \citep{kumar2018a, surdejetal24a}. In one year, it covers a total area of 115 square degrees, including 58 square degrees at high galactic latitude ($|b_{II}| > $ 30$^{\circ}$). The survey also covers a range of ecliptic latitudes, from 
$+6^\circ$ to  $+53^\circ$, enabling statistical studies of faint asteroid and TNO populations. 
Devasthal's longitude between Eastern Australia and Europe also makes it a strategic site for photometric studies of transients. 

Besides relative-low cost, the ILMT enjoys a number of benefits due to its zenith-pointing and TDI-mode of observing. Observations at the zenith 
have the lowest extinction, the darkest sky, and the best atmospheric seeing. In addition, the observation efficiency of the ILMT is very high ($\sim$ 93 \%). No time is wasted in telescope slewing and acquiring the field. The only time lost is that needed to write the image to disk, which requires about 70 seconds every 17 minutes. That is not a fundamental limitation, but a consequence of the camera control software. Other LMTs (NODO and LZT) have achieved 100\% observing efficiency. 

The TDI mode of operation results in very-low background variations. This is because every pixel of the image is an average over all pixels in a CCD column. For the ILMT detector, this reduces background variations by a factor of 64. The ILMT is therefore well suited to the detection and characterisation of objects with low surface brightness, such as galaxies and diffuse objects \citep{surdejetal24b,fuetal24}.  

Finally, there is the cost benefit of a LMT. The ILMT is estimated to have cost about 2 million Euros, which is more than an order of magnitude less than the cost of comparable conventional telescopes.
  
The participation of ARIES provides opportunities for follow-up observations of the transients detected by the ILMT. These are carried out in target-of-opportunity mode using the ARIES 1.3m DFOT and 3.6m DOT telescopes. Spectroscopy with the ARIES Devasthal Faint Object Spectrograph \& Camera (ADFOSC) instrument mounted on DOT (\citealt{omar2019}) is used for classification and more detailed study (see preliminary results for supernova SN 2023af in \citealt{kumaretal24}).

We have demonstrated that the ILMT performs as expected, achieving an image quality that approaches the seeing limit, and reaching limiting magnitudes near 22 in the g', r', and i' spectral bands in a single exposure (102 s). As observations continue, these limits grow increasingly faint in the coadded data. A transient pipeline has been established that efficiently identifies transients. Work is continuing to further develop the ILMT scientific programmes and data analysis techniques. 

\begin{acknowledgements}
The authors thank Ankit Bisht, Nikhil Dharkiya and other observing staff for their assistance at the 4m ILMT.  The team acknowledges the contributions of AMOS (Advanced Mechanical and Optical Systems), CSL (Centre Spatial de Li{\`e}ge), Socabelec (Jemeppe-sur-Sambre), Prof. Ram Sagar and ARIES's past and present scientific, engineering and administrative members in the realisation of the ILMT project. We are grateful to Serge Habraken and Jean-Pierre Swings  (University of Li\`ege) for their past  contributions to the project. JS wishes to thank the University of Li\`{e}ge, Service Public Wallonie and F.R.S.-FNRS (Belgium) for funding the construction of the ILMT. PH acknowledges financial support from the Natural Sciences and Engineering Research Council of Canada, RGPIN-2019-04369. PH, ST, AP-S and JS thank ARIES for hospitality during their visits to Devasthal. BA acknowledges the Council of Scientific $\&$ Industrial Research (CSIR) fellowship award (09/948(0005)/2020-EMR-I) for this work. MD acknowledges Innovation in Science Pursuit for Inspired Research (INSPIRE) fellowship award (DST/INSPIRE Fellowship/2020/IF200251) for this work. TA thanks Ministry of Higher Education, Science and Innovations of Uzbekistan (grant FZ-20200929344). JS, DB and KM acknowledge the assistance and support received from the Anusandhan National Research Foundation (ANRF, SERB-762 VAJRA Faculty Scheme, India).
\end{acknowledgements}

\bibliographystyle{aa}
\bibliography{main_surdej}

\end{document}